\documentclass[paper,notoc,nohyper]{JHEP}
\usepackage{epsfig}
\def\S{S_{\epsilon}}

\def\F{{\cal F}}

\def\bzero{\beta_0}

\def\Bfin{{\rm Box^6} (u, t) }
\def\Bubl{{\rm Bub}(s)}

\def\lib{{\rm Li}_2}
\def\lic{{\rm Li}_3}
\def\lid{{\rm Li}_4}
\def\Ls{L_s}
\def\lnx{L_x}
\def\lny{L_y}
\def\Poles{{\cal P}oles}
\def\Finite{{\cal F}inite}
\def\Libx{{\rm Li}_2(x)}
\def\Liby{{\rm Li}_2(y)}
\def\Licx{{\rm Li}_3(x)}
\def\Licy{{\rm Li}_3(y)}
\def\Lidx{{\rm Li}_4(x)}
\def\Lidy{{\rm Li}_4(y)}
\def\Lidz{{\rm Li}_4\Biggl(\frac{x-1}{x}\Biggr)}

\def\ust{\frac{u^2}{st}}
\def\ut{\frac{u}{s}}
\def\tos{\frac{t^2}{s^2}}
\def\sot{\frac{s^2}{t^2}}

\def\sut{\frac{s^2}{ut}}
\def\st{\frac{s}{u}}
\def\tou{\frac{t^2}{u^2}}
\def\uot{\frac{u^2}{t^2}}

\def\CA{C_A}
\def\CF{C_F}

\def\NF{N_F}

\def\A{{\cal A}}
\def\B{{\cal B}}
\renewcommand\O[1]{{\cal O}\left(#1\right)}
\def\as{\ensuremath{\alpha_{s}}}
\def\a0{\alpha_0}

\def\Re{\mathop{\rm Re}}

\def\beq{\begin{equation}}
\def\eeq{\end{equation}}

\def\beqn{\begin{eqnarray}}
\def\eeqn{\end{eqnarray}}

\def\lq{\left[}
\def\rq{\right]}

\def\({\left(}
\def\){\right)}

\def\ket#1{|{#1}\rangle}
\def\bra#1{\langle{#1}|}
\def\braket#1#2{\langle #1 |#2 \rangle}
\def\cm{{\cal M}}
\def\cmb{\overline{{\cal M}}}

\def\MSbar{$\overline{{\rm MS}}$}

\def\bom#1{{\mbox{\boldmath $#1$}}}

\def\fs{\(-\frac{\mu^2}{s}\)^\ep }
\def\ft{\(-\frac{\mu^2}{t}\)^\ep }
\def\fu{\(-\frac{\mu^2}{u}\)^\ep }
\def\fsd{\(-\frac{\mu^2}{s}\)^{2\ep} }
\def\ftd{\(-\frac{\mu^2}{t}\)^{2\ep} }
\def\fud{\(-\frac{\mu^2}{u}\)^{2\ep} }
\def \ep{\epsilon}

\newcommand{\SUNC}[1]{
\mbox{\parbox{3cm}{
\begin{picture}(3.5,1.4)
\thicklines
\put(0.5,0.7){\line(1,0){2}}
\put(1.5,0.7){\circle{1}}
\put(2.65,0.7){\makebox(0,0)[l]{$(#1)$}}
\end{picture}
}}
\hfill}

\newcommand{\GLASS}[1]{
\mbox{\parbox{3cm}{
\begin{picture}(3.5,1.4)
\thicklines
\put(0.5,0.7){\line(1,0){0.5}}
\put(3,0.7){\line(1,0){0.5}}
\put(1.5,0.7){\circle{1}}
\put(2.5,0.7){\circle{1}}
\put(3.65,0.7){\makebox(0,0)[l]{$(#1)$}}
\end{picture}
}}
\hfill}

\newcommand{\TRI}[1]{
\mbox{\parbox{3cm}{
\begin{picture}(3.5,1.4)
\thicklines
\put(0.5,0.7){\line(1,0){0.5}}
\put(1.5,1.2){\line(0,-1){1}}
\put(1.5,1.2){\line(1,0){1}}
\put(1.5,0.2){\line(1,0){1}}
\put(1.5,0.7){\circle{1}}
\put(2.65,0.7){\makebox(0,0)[l]{$(#1)$}}
\end{picture}
}}
\hfill}

\newcommand{\ABOX}[2]{
\mbox{\parbox{3cm}{
\begin{picture}(3.5,1.4)
\thicklines
\put(0.5,0.2){\line(1,0){2.4}}
\put(0.5,1.2){\line(1,0){2.4}}
\put(1,0.2){\line(0,1){1}}
\put(2,0.7){\circle{1}}
\put(3,0.7){\makebox(0,0)[l]{$(#1,#2)$}}
\end{picture}
}}
\hfill}

\newcommand{\CBOX}[2]{
\mbox{\parbox{3cm}{
\begin{picture}(3.5,1.4)
\thicklines
\put(0.5,0.2){\line(1,0){2.4}}
\put(1.2,0.2){\line(1,1){1}}
\put(0.5,1.2){\line(1,0){2.4}}
\put(1.2,0.2){\line(0,1){1}}
\put(2.2,0.2){\line(0,1){1}}
\put(3.05,0.7){\makebox(0,0)[l]{$(#1,#2)$}}
\end{picture}
}} 
\hfill}

\newcommand{\Pboxa}[2]{
\mbox{\parbox{3cm}{
\begin{picture}(3.5,1.4)
\thicklines
\put(0.5,0.2){\line(1,0){2.4}}
\put(1.7,0.2){\line(0,1){1}}
\put(0.5,1.2){\line(1,0){2.4}}
\put(1,0.2){\line(0,1){1}}
\put(2.4,0.2){\line(0,1){1}}
\put(3.05,0.7){\makebox(0,0)[l]{$(#1,#2)$}}
\end{picture}
}} 
\hfill}

\newcommand{\Pboxb}[2]{
\mbox{\parbox{3cm}{
\begin{picture}(3.5,1.4)
\thicklines
\put(0.5,0.2){\line(1,0){2.4}}
\put(1.7,0.2){\line(0,1){1}}
\put(0.5,1.2){\line(1,0){2.4}}
\put(1,0.2){\line(0,1){1}}
\put(2.4,0.2){\line(0,1){1}}
\put(1.35,0.7){\circle{0.5}}
\put(1.35,0.7){\makebox(0,0)[c]{$1$}}
\put(3.05,0.7){\makebox(0,0)[l]{$(#1,#2)$}}
\end{picture}
}} 
\hfill}

\newcommand{\Pboxc}[2]{
\mbox{\parbox{3cm}{
\begin{picture}(3.5,1.4)
\thicklines
\put(0.5,0.2){\line(1,0){2.4}}
\put(1.7,0.2){\line(0,1){1}}
\put(0.5,1.2){\line(1,0){2.4}}
\put(1,0.2){\line(0,1){1}}
\put(2.4,0.2){\line(0,1){1}}
\put(1.7,0.7){\circle*{0.2}}
\put(3.05,0.7){\makebox(0,0)[l]{$(#1,#2)$}}
\end{picture}
}} 
\hfill}

\newcommand{\Xboxa}[2]{
\mbox{\parbox{3cm}{
\begin{picture}(3.5,1.4)
\thicklines
\put(0.5,0.2){\line(1,0){2.4}}
\put(0.5,1.2){\line(1,0){2.4}}
\put(1,0.2){\line(0,1){1}}
\put(1.7,0.2){\line(1,1){1}}
\put(1.7,1.2){\line(1,-1){1}}
\put(3.05,0.7){\makebox(0,0)[l]{$(#1,#2)$}}
\end{picture}
}} 
\hfill}

\newcommand{\Xboxb}[2]{
\mbox{\parbox{3cm}{
\begin{picture}(3.5,1.4)
\thicklines
\put(0.5,0.2){\line(1,0){2.4}}
\put(1.7,0.2){\line(1,1){1}}
\put(0.5,1.2){\line(1,0){2.4}}
\put(1,0.2){\line(0,1){1}}
\put(1.7,1.2){\line(1,-1){1}}
\put(1,0.7){\circle*{0.2}}
\put(3.05,0.7){\makebox(0,0)[l]{$(#1,#2)$}}
\end{picture}
}} 
\hfill}

\title{\boldmath Two-loop QCD corrections to massless identical quark
scattering\footnote{Work supported in part by the UK Particle Physics and
Astronomy Research Council and by the EU Fourth Framework Programme
`Training and Mobility of Researchers', Network `Quantum Chromodynamics
and the Deep Structure of Elementary Particles',
contract FMRX-CT98-0194 (DG 12 - MIHT).
C.A. acknowledges
the financial support of the Greek government and
M.E.T. acknowledges financial support
from CONACyT and the CVCP. We thank
the British Council and German Academic Exchange Service for support
under ARC project 1050.
}}
\author{
C.~Anastasiou$^a$,
E.~W.~N.~Glover$^a$,
C.~Oleari$^b$ and M.~E.~Tejeda-Yeomans$^a$\\
$^a$Department of Physics, 
University of Durham, 
Durham DH1 3LE, 
England\\[1mm]
$^b$Department of Physics, 
University of Wisconsin,
1150 University Avenue\\
Madison WI 53706, 
U.S.A.\\[1mm]
E-mail: \email{Ch.Anastasiou@durham.ac.uk}, \email{E.W.N.Glover@durham.ac.uk},
\email{Oleari@pheno.physics.wisc.edu}, 
\email{M.E.Tejeda-Yeomans@durham.ac.uk}}
\abstract{
We present the  
two-loop virtual QCD corrections to the scattering of identical
massless quarks,
$q \bar q \to  q \bar q$, 
in conventional dimensional
regularisation and using the \MSbar\ scheme.  
The structure of the infrared divergences agrees with
that predicted by Catani while expressions for the finite
remainder are given for the 
$ q \bar q \to  q \bar q$ and
the $q q \to q q$ ($\bar q \bar q \to \bar q \bar q$) scattering processes
in terms of polylogarithms.   The results presented here form a vital
part of the next-to-next-to-leading order contribution to inclusive 
jet production in hadron colliders and will play a crucial
role in improving the theoretical prediction for jet cross 
sections in hadron-hadron collisions. }
\preprint{{DTP/00/70},{IPPP/00/08},{MADPH-00-1200},{hep-ph/0011094}}

\begin{document}

\section{Introduction}
\label{sec:intro}

Jet production at large transverse energies is a direct test of
parton-parton scattering processes in hadron-hadron collisions. At large jet
energy scales, the point-like nature of the partons can be probed down to
distance scales of  about $10^{-17}$~m by comparing data with QCD
predictions. Within the experimental and theoretical uncertainties,
data from the TEVATRON and CERN S$p\bar{p}$S generally show
good agreement with  the state-of-the-art theoretical next-to-leading order
$\O{\as^3}$ estimates based on massless parton-parton scattering
over a wide range of jet energies~\cite{EKS,jetrad}. 
It is anticipated that
the forthcoming Run
II starting at the TEVATRON in 2001   will yield a dramatic improvement  in
the quality of the data with increased statistics and improved detectors,
leading to a significant reduction  in both the statistical and systematic
errors. Subsequently, the start of data taking at the LHC will lead to a
much enlarged range of jet energies being probed.

It is a challenge to the physics community to improve the quality of the
theoretical predictions to a level that matches the improved experimental
accuracy. This may be achieved  by
including the  next-to-next-to-leading order QCD corrections
which  both reduces the renormalisation scale dependence and improves the
matching of the parton-level theoretical jet algorithm with the hadron-level
experimental jet algorithm. 

The full next-to-next-to-leading order prediction is a formidable task and 
requires a knowledge of
the two-loop $2 \to 2$ matrix elements as well as the contributions from the
one-loop  $2 \to 3$ and tree-level $2 \to 4$ processes.  
In the interesting
large-transverse-energy region, $E_T \gg m_{\rm quark}$, 
the quark masses may be
safely neglected and we therefore focus on the scattering of massless partons.
For processes involving up, down and strange quarks, which together with
processes involving gluons form the bulk of the cross section, 
this is certainly a reliable approximation.   
The contribution involving charm and bottom quarks 
is only a small part of the total since the parton densities for finding 
charm and bottom quarks inside the proton are relatively suppressed. 
We note that the existing next-to-leading order 
programs~\cite{EKS,jetrad} used to compare directly with the experimental 
jet data~\cite{CDF,D0} are based on
massless parton-parton scattering.  Helicity
amplitudes for the one-loop $2 \to 3$ parton sub-processes  have been
computed in~\cite{5g,3g2q,1g4q} while the amplitudes for the tree-level
$2\to 4$ processes are available in~\cite{6g,4g2q,2g4q,6q}.  The 
parton-density functions are also needed to next-to-next-to-leading order
accuracy. 
This requires knowledge of the three-loop splitting functions. At
present, the even moments of the splitting functions are known for the
flavour singlet and non-singlet structure functions $F_2$ and $F_L$ up to
$N=12$  while the odd moments up to $N=13$ are known for $F_3$
\cite{moms1,moms2}. The numerically small $\NF^2$ non-singlet contribution
is also known~\cite{Gra1}. Van Neerven and Vogt have provided accurate
parameterisations  of the splitting functions in $x$-space~\cite{NV,NVplb}
which are now starting to be implemented in the global analyses~\cite{MRS}.

The calculation of the two-loop amplitudes for the   $2 \to 2$ scattering of
light-like particles has proved more intractable due mainly to the difficulty
of evaluating the planar and non-planar double box graphs. Recently, however,
analytic expressions  for these basic scalar integrals  have been provided by
Smirnov~\cite{planarA} and Tausk~\cite{nonplanarA}  as series  in
$\ep=(4-D)/2$.  Algorithms for computing the associated tensor integrals have
also 
been provided~\cite{planarB} and~\cite{nonplanarB}, so that generic two-loop
massless  $2 \to 2$ processes can in principle be expressed in terms of a basis
set of known two-loop integrals.  Bern, Dixon and Kosower~\cite{bdk} were the
first to address such scattering processes and provided
analytic expressions for the maximal-helicity-violating two-loop amplitude for
$gg \to gg$. Subsequently,  Bern, Dixon and Ghinculov~\cite{BDG}  completed the
two-loop calculation of physical $2 \to 2$ scattering amplitudes for the QED 
processes $e^+e^- \to \mu^+\mu^-$ and $e^+e^- \to e^-e^+$. 

In an earlier paper~\cite{qqQQ}, we derived expressions for the two-loop
contribution to unlike quark scattering, $q \bar q \to q^\prime
\bar{q}^\prime $, 
as well as the crossed and time reversed processes.  The infrared pole
structure agreed with that predicted by Catani~\cite{catani} and we provided
explicit formulae for the finite parts in the $s$-, $t$- and $u$-channels in
terms of logarithms and polylogarithms. Matrix elements for the other 
parton-parton scattering processes remain to be evaluated.   
In this paper we extend the work of~\cite{qqQQ} to describe the case of
identical quark scattering.  We use the \MSbar\  renormalisation scheme
and conventional dimensional regularisation where all external particles
are treated in $D$ dimensions to provide dimensionally regularised and
renormalised analytic expressions  at the two-loop level for the  
scattering process
$$ 
q \bar q \to  q  \bar q,
$$ 
together with the time-reversed and crossed processes
\begin{eqnarray*} 
q + q&\to & q + q,\\ 
\bar{q} + \bar{q} &\to & \bar{q} +\bar{q}. 
\end{eqnarray*} 
As in the unlike quark case, we present analytic expressions for the infrared
pole structure, as well as explicit formulae for the finite remainder
decomposed 
according to powers of the number of colours $N$ and the number of light-quark
flavours $\NF$.
For the contributions most subleading in $N$,
there is an overlap
with the two-loop contribution to Bhabha scattering described in~\cite{BDG}
and the analytic expressions
presented here provide a useful check of some of their results.

Our paper is organised as follows.  We first establish our notation in 
Section~\ref{sec:notation}.  The results are collected in
Section~\ref{sec:results} where we provide analytic
expressions for the interference of the two-loop and tree-level amplitudes
as series expansions in $\ep$.  In Section~\ref{subsec:poles} we adopt the
notation of Catani~\cite{catani} to isolate the infrared singularity
structure of the two-loop amplitudes in the \MSbar\  scheme. We give
explicit formulae for the pole structure  obtained
by direct evaluation of the Feynman diagrams and show that it agrees with
the pole structure expected  on general grounds.   The finite $\O{\ep^0}$
remainder of the two-loop graphs is the main result of  our paper and
expressions appropriate for the $q \bar q \to  q\bar q $ and $qq \to qq$
($\bar q \bar q \to \bar q \bar q$) scattering processes are are given in
Section~\ref{subsec:finite}.  Finally Section~\ref{sec:conc} contains a brief
summary of our results.

\section{Notation} 
\label{sec:notation}

For calculational convenience, we treat all particles as incoming so that
\begin{equation}
\label{eq:proc}
q (p_1) + \bar q (p_2)  + q (p_3) + \bar{q}(p_4) \to 0,
\end{equation}
where the light-like momentum assignments are in parentheses and satisfy
$$
p_1^\mu+p_2^\mu+p_3^\mu+p_4^\mu = 0.
$$

As stated above,
we work in conventional dimensional regularisation  treating all external
states in $D$ dimensions.  We renormalise in the \MSbar\  scheme where
the bare coupling $\a0$ is related to the running coupling $\as \equiv \alpha_s(\mu^2)$  
at renormalisation scale $\mu$ via
\beq
\label{eq:alpha}
\a0 \,  \S = \as \,  \lq 1 - \frac{\beta_0}{\ep}  
\, \left(\frac{\as}{2\pi}\right) + \( \frac{\beta_0^2}{\ep^2} - \frac{\beta_1}{2\ep} \)  \, 
\left(\frac{\as}{2\pi}\right)^2
+\O{\as^3} \rq.
\eeq
In this expression
\beq
\S = (4 \pi)^\ep e^{-\ep \gamma},  \quad\quad \gamma=0.5772\ldots=
{\rm Euler\ constant}
\eeq
is the typical phase-space volume factor in $D=4-2\ep$ dimensions, and
$\beta_0, \beta_1$ are the first two coefficients of the QCD beta function for $\NF$ 
(massless) quark flavours
\beq
\label{betas}
\beta_0 = \frac{11 \CA - 4 T_R \NF}{6} \;\;, \;\; \;\;\;\;
\beta_1 = \frac{17 \CA^2 - 10 \CA T_R \NF - 6 \CF T_R \NF}{6} \;\;.
\eeq
For an $SU(N)$ gauge theory ($N$ is the number of colours)
\beq
\CF = \left(\frac{N^2-1}{2N}\right), \qquad \CA = N, \qquad T_R = \frac{1}{2}.
\eeq
The renormalised four point amplitude in the \MSbar\  scheme is thus
\beqn
\ket{\cm}&=& 4\pi \as \Biggl [ \left(\ket{\cm^{(0)}} -\ket{\cmb^{(0)}}\right)
+ \left(\frac{\as}{2\pi}\right) \left(\,\ket{\cm^{(1)}}-\ket{\cmb^{(1)}}\right)
\nonumber \\
&& \hspace{4cm}+ \left(\frac{\as}{2\pi}\right)^2\,
\left(\ket{ \cm^{(2)}}-\ket{ \cmb^{(2)}}\right) + \O{\as^3} \Biggr ],
\eeqn
where the $\ket{\cm^{(i)}}$ represents the colour-space vector describing the
$i$-loop amplitude for the $s$-channel graphs,
and the $t$-channel contribution
$\ket{\cmb^{(i)}}$ is obtained by exchanging the roles of
particles 2 and 4:
\beq
\ket{\cmb^{(i)}} = \ket{\cm^{(i)}} ( 2 \leftrightarrow 4).
\eeq
The dependence on both renormalisation scale $\mu$ and
renormalisation scheme is implicit.

We denote the squared amplitude summed over spins and colours by
\beqn
\braket{\cm}{\cm} &=& \sum |{\cal M}({q + \bar{q} \to  \bar{q} + q })|^2
\nonumber \\  
&=& \A(s,t,u) + \A(t,s,u) + \B(s,t,u),
\eeqn
where the Mandelstam variables are given by
\begin{equation}
s = (p_1+p_2)^2, \qquad t = (p_2+p_3)^2, \qquad u = (p_1+p_3)^2.
\end{equation}
The squared matrix elements for the $q q \to qq$ 
process are obtained by exchanging $s$ and $u$
\beq
\label{eq:AB}
\sum |{\cal M}({q + q \to  q + q })|^2  
= \A(u,t,s) + \A(t,u,s) + \B(u,t,s).
\eeq

The function $\A$  is related to the squared matrix elements for 
unlike quark scattering
\beqn
\A(s,t,u) &=& \sum |{\cal M}({q + \bar{q} \to  \bar{q}^\prime + q^\prime }
)|^2\\
\A(t,s,u) &=& \sum |{\cal M}({q + \bar{q}^\prime \to  \bar{q}^\prime +
q})|^2
\eeqn
while $\B(s,t,u)$ represents the interference between 
$s$-channel and $t$-channel graphs that is only present for identical quark
scattering.

The function $\A$ can be expanded perturbatively to yield
\beqn
\A(s,t,u) &=& 16\pi^2\as^2 \left[
 \A^4(s,t,u)+\left(\frac{\as}{2\pi}\right) \A^6(s,t,u)
 +\left(\frac{\as}{2\pi}\right)^2 \A^8(s,t,u) +
\O{\as^{3}}\right],\nonumber \\  
\eeqn
where
\beqn
\A^4(s,t,u) &=& \braket{\cm^{(0)}}{\cm^{(0)}} \equiv 2(N^2-1) \left(\frac{t^2+u^2}{s^2} - \epsilon
\right),\\
\A^6(s,t,u) &=& \left(
\braket{\cm^{(0)}}{\cm^{(1)}}+\braket{\cm^{(1)}}{\cm^{(0)}}\right),\\
\A^8(s,t,u) &=& \left( \braket{\cm^{(1)}}{\cm^{(1)}} +
\braket{\cm^{(0)}}{\cm^{(2)}}+\braket{\cm^{(2)}}{\cm^{(0)}}\right). 
\eeqn
Expressions for $\A^6$ are given in Ref.~\cite{ES} using dimensional
regularisation to isolate the infrared and ultraviolet singularities.  
Analytical formulae for the two-loop contribution to $\A^8$,
$\braket{\cm^{(0)}}{\cm^{(2)}}+\braket{\cm^{(2)}}{\cm^{(0)}}$,  
are given in Ref.~\cite{qqQQ}. 

Similarly, the expansion of $\B$ can be written
\beqn
\B(s,t,u) &=& 16\pi^2\as^2 \left[
 \B^4(s,t,u)+\left(\frac{\as}{2\pi}\right) \B^6(s,t,u)
 +\left(\frac{\as}{2\pi}\right)^2 \B^8(s,t,u) +
\O{\as^{3}}\right],\nonumber \\
\eeqn
where, in terms of the amplitudes, we have
\beqn
\B^4(s,t,u) &=& -\left(\braket{\cmb^{(0)}}{\cm^{(0)}}
 +\braket{\cm^{(0)}}{\cmb^{(0)}}\right) \nonumber \\
&\equiv& -4\left(\frac{N^2-1}{N}\right) (1-\epsilon)
\left(\frac{u^2}{st} + \epsilon\right),\\
\B^6(s,t,u) &=& -\left(\braket{\cmb^{(1)}}{\cm^{(0)}}
+\braket{\cm^{(0)}}{\cmb^{(1)}}
+\braket{\cmb^{(0)}}{\cm^{(1)}} 
+\braket{\cm^{(1)}}{\cmb^{(0)}}\right)\nonumber \\
&&\\
\B^8(s,t,u) &=& -\left(
\braket{\cmb^{(1)}}{\cm^{(1)}}
+\braket{\cm^{(1)}}{\cmb^{(1)}}\right.\nonumber \\
&& \hspace{0.5cm}
\left.
+\braket{\cmb^{(0)}}{\cm^{(2)}}
+\braket{\cm^{(2)}}{\cmb^{(0)}}
+\braket{\cm^{(0)}}{\cmb^{(2)}}
+\braket{\cmb^{(2)}}{\cm^{(0)}}
\right).\nonumber \\
\eeqn
As before, expressions  for $\B^6$ which are valid in conventional
dimensional regularisation  are given in Ref.~\cite{ES}.  Here, in order to
complete the calculation of the two-loop contribution to  quark-quark
scattering,  we concentrate on the next-to-next-to-leading order
contribution  $\B^8$ and in particular the interference of the two-loop and
tree graphs.

As in Ref.~\cite{qqQQ}, we use {\tt QGRAF}~\cite{QGRAF} to  produce the
two-loop Feynman diagrams to construct either  $\ket{\cm^{(2)}}$ or
$\ket{\cmb^{(2)}}$. We then project by $\bra{\cmb^{(0)}}$ or
$\bra{\cm^{(0)}}$ respectively and perform the summation over colours and
spins.  Finally, the  trace over the Dirac matrices is carried  out in $D$
dimensions using conventional dimensional regularisation. It is then
straightforward to identify the scalar and tensor integrals present  and
replace them with combinations of the basis set  of master integrals using
the  tensor reduction of two-loop integrals described in
\cite{planarB,nonplanarB,AGO3}, based on integration-by-parts~\cite{IBP} and 
Lorentz invariance~\cite{diffeq} identities.   The final result is  a
combination of master integrals in $D=4-2\epsilon$.   The basis set we
choose comprises 
\beqn
{\rm Sunset}(s) &=& \SUNC{s}\\
{\rm Glass}(s) &=&   \GLASS{s} \\
{\rm Tri}(s) &=&  \TRI{s}\\
{\rm Abox}(s,t) &=&
\ABOX{s}{t} \\
{\rm Cbox}(s,t) &=& \CBOX{s}{t} \\
{\rm Pbox_1}(s,t) &=& \Pboxa{s}{t} \\
{\rm Xbox_1}(s,t) &=& \Xboxa{s}{t} \\
{\rm Xbox_2}(s,t) &=& \Xboxb{s}{t}
\eeqn
and\footnote{Reference~\cite{planarB} describes the procedure for reducing the tensor integrals down to a basis
involving the planar box integral 
$$
{\rm Pbox_2}(s,t) = \Pboxc{s}{t}, 
$$ 
where the  blob on the middle propagator represents an additional power of
that propagator, 
and provides a series expansion for ${\rm Pbox}_2$ to $\O{\epsilon^0}$.  However, as was
pointed out in~\cite{bastei}, knowledge of ${\rm Pbox_1}$ and ${\rm Pbox_2}$ to
$\O{\epsilon^0}$ is not sufficient to determine all tensor loop integrals to the same order. 
Series expansions for ${\rm Pbox_3}$ are relatively compact and straightforward to 
obtain and
are detailed in~\cite{bastei3,bastei2}.
${\rm Pbox_2}$ can therefore be eliminated in favor of ${\rm Pbox_3}$.
We note that this choice is not unique.   Bern et al.~\cite{BDG} choose to use
the ${\rm Pbox_1}$
and ${\rm Pbox_2}$ basis, but with the integrals evaluated in $D=6-2\epsilon$
dimensions where they are both infrared and ultraviolet finite. 
}
\beq 
{\rm Pbox_3}(s,t) = \Pboxb{s}{t}, 
\eeq
where  
\begin{picture}(1,1)
\thicklines
\put(0.5,0.2){\circle{0.5}}
\put(0.5,0.2){\makebox(0,0)[c]{$1$}}
\end{picture}
represents the planar box integral with one irreducible numerator associated with the left
loop.
The expansion in $\ep$ for all the non-trivial master integrals 
can be found 
in~\cite{planarA,nonplanarA,planarB,nonplanarB,AGO2,AGO3,xtri,bastei3,bastei2}.   

\section{Results}
\label{sec:results}
In this section, we give explicit formulae for the $\epsilon$-expansion of the
two-loop contribution to the next-to-next-to-leading order term $\B^8(s,t,u)$.  
To distinguish between
the genuine two-loop contribution and the squared
one-loop part, we decompose $\B^8$ as
\begin{equation}
\B^8 = \B^{8~(2\times 0)} + \B^{8~(1\times 1)}.
\end{equation}
The one-loop-square contribution $\B^{8~(1\times 1)}$ is vital in determining 
$\B^8$ but is relatively straightforward to
obtain. 
For the remainder of this paper we concentrate on the technically more
complicated two-loop contribution $\B^{8~(2\times 0)}$.

As in Ref.~\cite{qqQQ}, we divide the two-loop contributions into two
classes: those that multiply
poles in the dimensional regularisation parameter $\ep$ and those that are finite
as $\ep \to 0$
\beq
\B^{8~(2\times 0)}(s,t,u)
 = \Poles+\Finite.
\eeq 
$\Poles$ contains infrared singularities that will be  analytically
canceled by the infrared singularities occurring in radiative processes of the
same order (ultraviolet divergences are removed by renormalisation).

\subsection{Infrared Pole Structure}
\label{subsec:poles}

Catani has made predictions for the singular
infrared behaviour of two-loop amplitudes. 
Following the procedure advocated
in~\cite{catani}, we find that the pole structure in the \MSbar\ 
scheme can be written as
\beqn
\label{eq:poles}
\Poles = -2 \Re \Biggl [&&  
\frac{1}{2}\bra{\cmb^{(0)}} {\bom I}^{(1)}(\ep){\bom I}^{(1)}(\ep) \ket{\cm^{(0)}}
  -\frac{\beta_0}{\epsilon}  
\,\bra{\cmb^{(0)}} {\bom I}^{(1)}(\ep)  \ket{\cm^{(0)}}
 \nonumber\\
&& 
+\,\bra{\cmb^{(0)}} {\bom I}^{(1)}(\ep)  \ket{\cm^{(1) {\rm fin}}}
 \nonumber\\
&& 
+
e^{-\ep \gamma } \frac{ \Gamma(1-2\ep)}{\Gamma(1-\ep)} 
\left(\frac{\beta_0}{\epsilon} + K\right)
\bra{\cmb^{(0)}} {\bom I}^{(1)}(2\ep) \ket{\cm^{(0)}}\nonumber \\
&&+ \, \bra{\cmb^{(0)}}{\bom H}^{(2)}(\ep)\ket{\cm^{(0)}}  + ( s \leftrightarrow t)\Biggr ],
\eeqn
where the constant $K$ is
\beq
K = \left( \frac{67}{18} - \frac{\pi^2}{6} \right) \CA - \frac{10}{9} T_R
\NF.
\eeq 
In Eq.~(\ref{eq:poles}), the symmetrisation under $s$ and $t$ exchange 
represents the additional effect of the
$s$-channel tree graph interfering with the $t$-channel two-loop graphs.

The colour algebra is straightforward and we find that the $s$-$t$ symmetric
contributions proportional to 
\beq
\braket{\cmb^{(0)}}{\cm^{(0)}} = 2 \left(\frac{N^2-1}{N}\right)
(1-\epsilon)\left(\frac{u^2}{st}+\epsilon \right),
\eeq
are given by
\beqn
\lefteqn{\bra{\cmb^{(0)}}{\bom I}^{(1)}(\ep)\ket{\cm^{(0)}} =
\braket{\cmb^{(0)}}{\cm^{(0)}}}\nonumber\\
&&\times\frac{e^{\ep \gamma}}{\Gamma(1-\ep)}
 \( \frac{1}{\ep^2} + \frac{3}{2 \ep}\) 
\Bigg[
 \frac{1}{N} \fs 
+\frac{1}{N} \ft 
-\frac{N^2+1}{N} \fu
\Bigg]   
\\ \nonumber \\ \nonumber \\
\lefteqn{\bra{\cmb^{(0)}}{\bom I}^{(1)}(\ep){\bom I}^{(1)}(\ep)\ket{\cm^{(0)}} 
=\braket{\cmb^{(0)}}{\cm^{(0)}}}\nonumber\\
&&\times\frac{e^{2\ep \gamma}}{\Gamma(1-\ep)^2}
 \( \frac{1}{\ep^2} + \frac{3}{2 \ep}\)^2 
 \Bigg\{
 \frac{N^4-3 N^2-2}{N^2} \fu\left[\fs+\ft\right] \nonumber \\
 && \hspace{1cm}
 + \frac{3N^2+1}{N^2} \fud
 - \frac{(N^2-2)(N^2+1)}{N^2}\fs \ft  
\nonumber\\
&&{}\hspace{1cm}
+ \frac{1}{N^2} \fsd 
+ \frac{1}{N^2} \ftd\Bigg\}
\eeqn
and
\beqn
\label{eq:htwo}
\lefteqn{\bra{\cmb^{(0)}}{\bom H}^{(2)}(\ep)\ket{\cm^{(0)}} = \braket{\cmb^{(0)}}{\cm^{(0)}} }\nonumber \\ 
&&\hspace{1cm}\times\frac{e^{\ep \gamma}}{2\ep\Gamma(1-\ep)} H^{(2)} 
\Bigg[\fsd + \ftd - \fud 
\Bigg],
\eeqn
where the constant $H^{(2)}$ is
\beq
H^{(2)} =  \left [\frac{1}{4}\gamma_{(1)}+3\CF K + \frac{5}{2}\zeta_2\beta_0\CF -
\frac{28}{9}\beta_0 \CF - \left(\frac{16}{9}-7\zeta_3\right)\CF\CA \right ].
\eeq
Here $\zeta_n$ is the Riemann Zeta function, $\zeta_2 = \pi^2/6$, 
$\zeta_3 = 1.202056\ldots$  
and
\beq
\gamma_{(1)}=\(-3+24\zeta_2-48\zeta_3\)\CF^2
+\left(-\frac{17}{3}-\frac{88}{3}\zeta_2+24\zeta_3\right)\CF\CA
+\left(\frac{4}{3}+\frac{32}{3}\zeta_2\right)\CF T_R\NF.
\eeq
The square bracket in 
Eq.~(\ref{eq:htwo}) is a guess 
simply motivated by summing over the antennae present
in the quark-quark scattering process
and on
dimensional grounds.  Different choices affect only the finite remainder.

The bracket of ${\bom I}^{(1)}$ between the $t$-channel tree graph and the
finite part of the $s$-channel one-loop graphs is not symmetric under the
exchange of $s$ and $t$ and is given by 
\beqn
\label{eq:I1M1}
\lefteqn{\bra{\cmb^{(0)}}{\bom I}^{(1)}(\ep)\ket{\cm^{(1) {\rm fin}}} 
=\frac{e^{\ep \gamma}}{\Gamma(1-\ep)}
 \( \frac{1}{\ep^2} + \frac{3}{2 \ep}\) }\nonumber\\
&\times&  \Bigg\{ 
\left[ {1\over N}\fs +{1 \over N}
\ft - \frac{N^2+1}{N}  \fu  \right] {\cal F}_1(s,t,u)
\nonumber \\
&&   
+\left[ \frac{N^2-1}{N} \fs - {1 \over N}\ft +{1 \over N}\fu
\right] (N^2-1) \,{\cal F}_2(s,t,u)
\Bigg\}.\nonumber \\
\eeqn
The functions $\F_1$ and $\F_2$ appearing in Eq.~(\ref{eq:I1M1}) 
are finite and are given by
\begin{eqnarray}
{\cal F}_1(s,t,u) &=&
\frac{N^2-1}{2N^2} \left[ \(N^2-2\) f_1(s,t,u) +2f_2(s,t,u) \right]
\nonumber \\
&-&  
\frac{1}{2\ep \,(3-2\ep)} \left[
\frac{N^2-1}{N}
\(6-7\ep-2\ep^2\)-\frac{1}{N} \(10\ep^2-4\ep^3\) 
\right] \Bubl
 \braket{\cmb^{(0)}}{\cm^{(0)}}
\nonumber \\
&-&
\frac{e^{\ep \gamma}}{ \Gamma(1-\ep)} 
\left(\frac{1}{\ep^2}+\frac{3}{2 \ep} \right)
\nonumber \\
&\times&  
 \left[ 
\frac{1}{N} \fs 
-\frac{2}{N} \fu
-\frac{N^2-2}{N}\ft
 \right] \braket{\cmb^{(0)}}{\cm^{(0)}}
\nonumber \\
&-&  
\bzero \left[\frac{1}{\ep}-\frac{3 (1-\ep)}{3-2 \ep} \Bubl \right] \braket{\cmb^{(0)}}{\cm^{(0)}}
\nonumber \\
\end{eqnarray}
and
\begin{eqnarray}
{\cal F}_2(s,t,u) &=&
\frac{N^2-1}{2N^2} \Big[ f_1(s,t,u) - f_2(s,t,u) \Big] 
\nonumber \\
&&-\frac{e^{\ep \gamma}}{\Gamma(1-\ep)} 
\left(\frac{1}{\ep^2}+\frac{3}{ 2\ep} \right) 
\left[ {1 \over N}\fu -{1 \over N}\ft \right]  \braket{\cmb^{(0)}}{\cm^{(0)}}
\nonumber \\
&& 
\end{eqnarray}
with 
\begin{eqnarray}
f_1(s,t,u) &=& \frac{2 u}{s t} (1-2 \ep)  \left[ 
u^2+t^2 - 2 \ep \( t^2+s^2\) +\ep^2 s^2
\right] {\rm Box}^6(s, t)
\nonumber \\
&&+\frac{2}{ s t} \left[ 2 u^2-\ep \(5 s^2 +6 t^2+9 s t\)+
\(2 s^2 +4 t^2+s t\) \ep^2 \right. 
\nonumber \\
&& \hspace{1cm}\left. + \(s^2+3 s t \) \ep^3 -s t \ep^4 
  \right] \left[ \frac{{\rm Bub}(s) - {\rm Bub}(t) }{\ep}\right],\\
f_2(s,t,u) &=& \frac{2 }{s } (1-2 \ep)  \left[ 
2 u^2- \ep \( t^2+s^2+u^2\) +3 \ep^2  s^2
 +s^2 \ep^3 \right] {\rm Box}^6(s, u)
\nonumber \\
&&+\frac{2}{ s t} \left[ 2 u^2-\ep \(6 s^2 +6 t^2+10 s t\)+
\(3 s^2 +4 t^2+3 s t\) \ep^2 \right. 
\nonumber \\
&&\hspace{1cm} \left. + \(s^2+2 s t \) \ep^3 -s t \ep^4 
  \right] \left[\frac{ {\rm Bub}(s) - {\rm Bub}(u)}{\ep} \right].
\end{eqnarray}
These expressions are valid in all kinematic regions.  However, to evaluate the pole
structure in a particular region, they must be expanded as a series 
in $\epsilon$. We note that in Eq.~(\ref{eq:poles}) these functions are
multiplied by poles in $\epsilon$ and must therefore be expanded through to
$\O{\epsilon^2}$. In the physical region $ u < 0$, $t < 0$, $\Bfin$ has no imaginary part 
and is given by~\cite{BDG}
\begin{eqnarray}
\Bfin &=& \frac{ e^{\ep\gamma}
\Gamma  \left(  1+\epsilon \right)  \Gamma  
\left( 1-\epsilon \right) ^2 
 }{ 2s\Gamma  \left( 1-2 \epsilon  \right)   \left( 1-2 \epsilon  \right) } 
 \left(\frac{\mu^2}{s}\right)^{\ep}
  \Biggl [
 \frac{1}{2}\(\(\lnx-\lny\)^2+\pi^2 \)\nonumber \\
&& 
 +2\ep \left(
 \Licx-\lnx\Libx-\frac{1}{3}\lnx^3-\frac{\pi^2}{2}\lnx \right)
\nonumber \\
&& 
-2\ep^2\Biggl(
\Lidx+\lny\Licx-\frac{1}{2}\lnx^2\Libx-\frac{1}{8}\lnx^4-\frac{1}{6}\lnx^3\lny+\frac{1}{4}\lnx^2\lny^2\nonumber
\\
&&-\frac{\pi^2}{4}\lnx^2-\frac{\pi^2}{3}\lnx\lny-\frac{\pi^4}{45}\Biggr)
+ ( u \leftrightarrow t) \Biggr ] + \O{\ep^3},
\end{eqnarray}
where $x = -t/s$, $\lnx=\log(x)$ and $\lny=\log(1-x)$ and the 
polylogarithms ${\rm Li}_n(z)$ are defined by
\begin{eqnarray}
 {\rm Li}_n(z) &=& \int_0^z \frac{dt}{t} {\rm Li}_{n-1}(t) \qquad {\rm ~for~}
 n=2,3,4\\
 {\rm Li}_2(z) &=& -\int_0^z \frac{dt}{t} \log(1-t).
\end{eqnarray} 
Analytic continuation to other kinematic regions is obtained 
using the inversion formulae for the arguments of 
the polylogarithms (see for example~\cite{AGO3}) when $x > 1$
\begin{eqnarray}
\lib(x + i0) &=& -\lib\left(\frac{1}{x}\right) -\frac{1}{2}
 \log^2 (x) +\frac{\pi^2}{3} + i \pi \log (x) \nonumber \\
\lic(x + i0) &=& \lic\left(\frac{1}{x}\right) -\frac{1}{6}\log^3(x) +
\frac{\pi^2}{3}  \log(x) +  \frac{i\pi}{2} \log^2(x) \nonumber \\ 
\lid(x + i0)  &=& - \lid\left(\frac{1}{x}\right)
         -  \frac{1}{24}\log^4(x)  + \frac{\pi^2}{6} \log^2(x)
        + \frac{\pi^4}{45} +
        \frac{i\pi}{6}\log^3(x) .
\end{eqnarray}
Finally, the one-loop bubble integral in $D=4-2 \epsilon$ dimensions 
is given by
\begin{equation} 
 \Bubl =\frac{ e^{\ep\gamma}\Gamma  \left(  1+\epsilon \right)  \Gamma  \left( 1-\epsilon \right) ^2 
 }{ \Gamma  \left( 2-2 \epsilon  \right)  \epsilon   } \fs.
\end{equation}

The leading infrared singularity is $\O{1/\ep^4}$ and it is a very stringent
check on the reliability of our calculation that the pole structure obtained by
computing the Feynman diagrams agrees with that anticipated by Catani through
to $\O{1/\ep}$.   We therefore construct the finite remainder by subtracting
Eq.~(\ref{eq:poles}) from the full result.

\subsection{Finite contributions}
\label{subsec:finite}

In this subsection, we give explicit expressions for the finite two-loop
contribution to $\B^8$, $\Finite$, which is given by
\beq
\Finite = -2 \Re \left(
\braket{\cm^{(0)}}{\cmb^{(2)\, {\rm fin}}}
+\braket{\cmb^{(0)}}{\cm^{(2)\, {\rm fin}}}\right).
\eeq

The identical-quark processes probed in  
high-energy hadron-hadron collisions are the mixed
$s$- and $t$-channel process
$$
q + \bar{q} \to  \bar{q} + q,
$$
controlled by $\B(s,t,u)$ (as well as the distinct quark 
matrix elements $\A(s,t,u)$ and $\A(t,s,u)$ as indicated in Eq.~(\ref{eq:AB})),
and the mixed 
$t$- and $u$-channel processes
\begin{eqnarray*}
q + q&\to&  q + q,\\
\bar q + \bar q&\to&  \bar q + \bar q,
\end{eqnarray*}
which are determined by the 
$\B(t,s,u)$. We need to be able to evaluate the
finite parts for each of these processes.
Of course, the analytic expressions for different channels 
are related by crossing
symmetry.   However, the master crossed boxes have cuts in all three channels
yielding complex parts in all physical regions.   The analytic
continuation is therefore rather involved and prone to error.   We
therefore choose to give expressions 
describing $\B^8(s,t,u)$ and $\B^8(t,s,u)$ which are directly 
valid in the physical region $s > 0$ and
$u, t < 0$, and are given in terms of logarithms and polylogarithms that
have no imaginary parts.

Using the standard polylogarithm identities~\cite{kolbig} 
we retain the polylogarithms with arguments $x$, $1-x$ and
$(x-1)/x$, where
\begin{equation}
\label{eq:xydef}
x = -\frac{t}{s}, \qquad y = -\frac{u}{s} = 1-x, \qquad \frac{x-1}{x} =
-\frac{u}{t}.
\end{equation}
For convenience, we also introduce the following logarithms
\begin{equation}
\lnx = \log\left(\frac{-t}{s}\right),
\qquad \lny = \log\left(\frac{-u}{s}\right),
\qquad \Ls = \log\left(\frac{s}{\mu^2}\right),
\end{equation}
where $\mu$ is the renormalisation scale.
The common choice $\mu^2 = s$ corresponds to setting $\Ls = 0$.

For each channel, we choose to present our results by grouping terms 
according to the
power of the number of colours $N$ and the number 
of light quarks $\NF$, so that in channel $c$ 
\begin{equation}
\label{eq:zi}
\Finite_c =
 2 \left(\frac{N^2-1}{N}\right)
 \left(N^2 A_c + B_c  + \frac{1}{N^2} C_c
+ N \NF D_c   +\frac{\NF}{N} E_c  + \NF^2 F_c\right).
\end{equation}
Here $c = st$ ($ut$) to denote the mixed $s$- and $t$-channel ($u$- and
$t$-channel) processes respectively.

\subsubsection{The process $q \bar q \to \bar q q$}
\label{subsec:stex}
We first give expressions for the mixed $s$-channel and $t$-channel
annihilation process,
$q \bar q \to \bar q q$.
We find that
\begin{eqnarray}
A_{st}&=&{ }\Biggl [{2}\,{\Lidy}-{2}\,{\Lidx}+{2}\,{\Lidz}+{}\Biggl ({}-{2}\,{\lnx}+{12}\Biggr ){}\,{\Licy}+{4}\,{\lny}\,{\lnx}\,{\Liby}  \nonumber \\
&& +{}\Biggl ({}-{23\over 3}-{2}\,{\lnx}+{4}\,{\lny}\Biggr ){}\,{\Licx}+{}\Biggl ({23\over 3}\,{\lnx}+{12}\,{\lny}+{2}\,{\lnx^2}+{5\over 3}\,{\pi^2}\Biggr ){}\,{\Libx} \nonumber \\
&& -{121\over 9}\,{\Ls^2}+{}\Biggl ({11\over 3}\,{\lnx^2}+{}\Biggl ({}-{22\over 9}-{22\over 3}\,{\lny}\Biggr ){}\,{\lnx}+{11\over 3}\,{\pi^2}+{22\over 3}\,{\lny^2}-{22}\,{\lny}+{592\over 27}\Biggr ){}\,{\Ls} \nonumber \\
&& -{1\over 6}\,{\lnx^4}+{}\Biggl ({14\over 9}+{5\over 3}\,{\lny}\Biggr ){}\,{\lnx^3}+{}\Biggl ({}-{11\over 12}\,{\pi^2}+{\lny^2}-{31\over 6}+{13\over 12}\,{\lny}\Biggr ){}\,{\lnx^2} \nonumber \\
&& +{}\Biggl ({1\over 3}\,{\lny^3}+{6}\,{\lny^2}+{}\Biggl ({8\over 3}\,{\pi^2}+{8\over 9}\Biggr ){}\,{\lny}+{89\over 36}\,{\pi^2}-{6}\,{\zeta_3}+{695\over 216}\Biggr ){}\,{\lnx} \nonumber \\
&& -{1\over 6}\,{\lny^4}+{22\over 9}\,{\lny^3}+{}\Biggl ({}-{169\over 18}+{1\over 6}\,{\pi^2}\Biggr ){}\,{\lny^2}+{}\Biggl ({61\over 18}\,{\pi^2}+{12}\,{\zeta_3}+{1673\over 108}\Biggr ){}\,{\lny} \nonumber \\
&& -{347\over 18}\,{\zeta_3}-{121\over 360}\,{\pi^4}-{23213\over 1296}-{8\over 3}\,{\pi^2}\Biggr ]{}\,{\ust} \nonumber \\
&& +{}\Biggl [{}-{4}\,{\Lidx}+{24}\,{\Licy}+{}\Biggl ({2}\,{\lnx}+{12}\Biggr ){}\,{\Licx} +{}\Biggl ({}-{2\over 3}\,{\pi^2}+{24}\,{\lny}-{12}\,{\lnx}\Biggr ){}\,{\Libx}\nonumber \\
&& +{1\over 12}\,{\lnx^4}-{19\over 12}\,{\lnx^3}+{}\Biggl ({}-{5\over 2}+{1\over 3}\,{\pi^2}\Biggr ){}\,{\lnx^2}+{}\Biggl ({}-{2}\,{\zeta_3}-{29\over 6}\,{\pi^2}+{12}\,{\lny^2}+{5}\,{\lny}\Biggr ){}\,{\lnx} \nonumber \\
&& +{7\over 45}\,{\pi^4}-{5\over 2}\,{\pi^2}-{12}\,{\zeta_3}-{4}\,{\lny}\,{\pi^2}\Biggr ]{}\,{\ut}+{}\Biggl [{3}\,{\lnx^2}+{3}\,{\pi^2}+{3}\,{\lny^2}-{6}\,{\lnx}\,{\lny}\Biggr ]{}\,{\tos}+{}\Biggl [{3}\,{\lny^2}\Biggr ]{}\,{\sot} \nonumber \\
&& -{32}\,{\Lidy}-{32}\,{\Lidz}+{8}\,{\lnx}\,{\Licy}+{}\Biggl ({2}-{28}\,{\lny}+{18}\,{\lnx}\Biggr ){}\,{\Licx} \nonumber \\
&& +{}\Biggl ({}-{2}\,{\lnx^2}+{}\Biggl ({}-{2}-{24}\,{\lny}\Biggr ){}\,{\lnx}-{2}\,{\pi^2}\Biggr ){}\,{\Libx}-{28}\,{\lny}\,{\lnx}\,{\Liby}-{11\over 12}\,{\lnx^4} \nonumber \\
&& +{}\Biggl ({}-{7\over 12}+{14\over 3}\,{\lny}\Biggr ){}\,{\lnx^3}+{}\Biggl ({}-{32}\,{\lny^2}+{1\over 2}\,{\lny}+{2}+{1\over 2}\,{\pi^2}\Biggr ){}\,{\lnx^2} \nonumber \\
&& +{}\Biggl ({}\Biggl ({6}+{20\over 3}\,{\pi^2}\Biggr ){}\,{\lny}-{3\over 2}\,{\pi^2}-{18}\,{\zeta_3}\Biggr ){}\,{\lnx}-{2}\,{\zeta_3}-{3}\,{\pi^2}-{6}\,{\lny^2}+{28}\,{\lny}\,{\zeta_3}+{1\over 3}\,{\pi^4},
\end{eqnarray}
\begin{eqnarray}
B_{st}&=&{ }\Biggl [{}-{8}\,{\Lidy}-{3}\,{\Lidx}-{8}\,{\Lidz}+{8}\,{\lnx}\,{\Licy}+{}\Biggl ({}-{6}-{12}\,{\lny}+{12}\,{\lnx}\Biggr ){}\,{\Licx} \nonumber \\
&& +{}\Biggl ({}-{6}\,{\pi^2}+{6}\,{\lnx}-{13\over 2}\,{\lnx^2}\Biggr ){}\,{\Libx}-{12}\,{\lny}\,{\lnx}\,{\Liby} \nonumber \\
&& +{}\Biggl ({}-{11\over 6}\,{\lnx^2}+{}\Biggl ({}-{22\over 3}\,{\lny}+{22\over 3}\Biggr ){}\,{\lnx}-{22}\,{\lny}+{11\over 3}\,{\pi^2}+{22\over 3}\,{\lny^2}+{176\over 3}\Biggr ){}\,{\Ls} \nonumber \\
&& -{7\over 24}\,{\lnx^4}-{7\over 9}\,{\lnx^3}+{}\Biggl ({}-{17\over 2}\,{\lny^2}+{7\over 2}\,{\lny}-{19\over 36}-{1\over 6}\,{\pi^2}\Biggr ){}\,{\lnx^2} \nonumber \\
&& +{}\Biggl ({\lny^3}-{27\over 2}\,{\lny^2}+{}\Biggl ({}-{3}\,{\pi^2}+{251\over 9}\Biggr ){}\,{\lny}+{181\over 9}-{5\over 6}\,{\pi^2}-{12}\,{\zeta_3}\Biggr ){}\,{\lnx} \nonumber \\
&& -{1\over 2}\,{\lny^4}+{103\over 9}\,{\lny^3}+{}\Biggl ({}-{242\over 9}+{5\over 2}\,{\pi^2}\Biggr ){}\,{\lny^2}+{}\Biggl ({12}\,{\zeta_3}+{98\over 3}+{127\over 18}\,{\pi^2}\Biggr ){}\,{\lny} \nonumber \\
&& +{581\over 18}\,{\zeta_3}-{31\over 360}\,{\pi^4}-{124\over 9}\,{\pi^2}-{30659\over 324}\Biggr ]{}\,{\ust} \nonumber \\
&& +{}\Biggl [{}-{6}\,{\Lidx}+{4}\,{\lnx}\,{\Licx}-{\lnx^2}\,{\Libx}-{22\over 3}\,{\lnx}\,{\Ls}-{1\over 24}\,{\lnx^4}-{5\over 18}\,{\lnx^3}-{47\over 3}\,{\lnx^2} \nonumber \\
&& +{}\Biggl ({24}\,{\lny}+{2\over 9}\,{\pi^2}+{128\over 9}-{4}\,{\zeta_3}\Biggr ){}\,{\lnx}+{1\over 15}\,{\pi^4}-{47\over 3}\,{\pi^2}\Biggr ]{}\,{\ut} \nonumber \\
&& +{}\Biggl [{}-{8}\,{\lnx}\,{\lny}+{4}\,{\lnx^2}+{4}\,{\pi^2}+{4}\,{\lny^2}\Biggr ]{}\,{\tos}+{}\Biggl [{}{4}\,{\lny^2}\Biggr ]{}\,{\sot} \nonumber \\
&& +{16}\,{\Lidy}+{16}\,{\Lidz}-{16}\,{\lnx}\,{\Licy}+{}\Biggl ({}-{12}\,{\lnx}+{8}\,{\lny}+{2}\Biggr ){}\,{\Licx} \nonumber \\
&& +{}\Biggl ({4}\,{\lnx^2}+{4}\,{\pi^2}-{2}\,{\lnx}\Biggr ){}\,{\Libx}+{8}\,{\lny}\,{\lnx}\,{\Liby}+{11\over 3}\,{\lnx^2}\,{\Ls}+{5\over 8}\,{\lnx^4}+{}\Biggl ({2}-{4\over 3}\,{\lny}\Biggr ){}\,{\lnx^3} \nonumber \\
&& +{}\Biggl ({}-{\lny}+{4}\,{\lny^2}-{163\over 9}\Biggr ){}\,{\lnx^2}+{}\Biggl ({}\Biggl ({8\over 3}\,{\pi^2}+{8}\Biggr ){}\,{\lny}+{11\over 3}\,{\pi^2}+{12}\,{\zeta_3}\Biggr ){}\,{\lnx} \nonumber \\
&& -{2}\,{\zeta_3}-{2\over 3}\,{\pi^4}-{4}\,{\pi^2}-{8}\,{\lny}\,{\zeta_3}-{8}\,{\lny^2},
\end{eqnarray}
\begin{eqnarray}
C_{st}&=&{ }\Biggl [{}-{2}\,{\Lidy}-{5}\,{\Lidx}-{2}\,{\Lidz}+{2}\,{\lnx}\,{\Licy}+{}\Biggl ({1}+{6}\,{\lnx}-{4}\,{\lny}\Biggr ){}\,{\Licx} \nonumber \\
&& +{}\Biggl ({}-{7\over 3}\,{\pi^2}-{5\over 2}\,{\lnx^2}-{\lnx}\Biggr ){}\,{\Libx}-{4}\,{\lny}\,{\lnx}\,{\Liby}-{1\over 8}\,{\lnx^4}+{}\Biggl ({5\over 12}-{1\over 3}\,{\lny}\Biggr ){}\,{\lnx^3}+{3}\,{\lny^3} \nonumber \\
&& +{}\Biggl ({17\over 12}\,{\pi^2}+{5\over 4}-{3\over 4}\,{\lny}-{5\over 2}\,{\lny^2}\Biggr ){}\,{\lnx^2}+{}\Biggl ({}-{9\over 2}\,{\lny^2}+{}\Biggl ({}-{5\over 3}\,{\pi^2}+{13}\Biggr ){}\,{\lny}+{8\over 3}\,{\pi^2}-{45\over 8}\Biggr ){}\,{\lnx} \nonumber \\
&& +{}\Biggl ({}-{21\over 2}+{4\over 3}\,{\pi^2}\Biggr ){}\,{\lny^2}+{}\Biggl ({93\over 4}+{13\over 6}\,{\pi^2}-{8}\,{\zeta_3}\Biggr ){}\,{\lny}+{19}\,{\zeta_3}-{31\over 6}\,{\pi^2}-{511\over 16}-{1\over 90}\,{\pi^4}\Biggr ]{}\,{\ust} \nonumber \\
&& +{}\Biggl [{}-{10}\,{\Lidx}+{6}\,{\lnx}\,{\Licx}+{}\Biggl ({}-{\lnx^2}-{2\over 3}\,{\pi^2}\Biggr ){}\,{\Libx}+{1\over 24}\,{\lnx^4}-{13\over 12}\,{\lnx^3} \nonumber \\
&& +{}\Biggl ({}-{5\over 2}+{1\over 3}\,{\pi^2}\Biggr ){}\,{\lnx^2}+{}\Biggl ({5}\,{\lny}+{5\over 2}\,{\pi^2}-{6}\,{\zeta_3}+{12}\Biggr ){}\,{\lnx}+{2\over 9}\,{\pi^4}-{5\over 2}\,{\pi^2}\Biggr ]{}\,{\ut} \nonumber \\
&& +{}\Biggl [{\lnx^2}-{2}\,{\lnx}\,{\lny}+{\pi^2}+{\lny^2}\Biggr ]{}\,{\tos}+{}\Biggl [{\lny^2}\Biggr ]{}\,{\sot} \nonumber \\
&& +{8}\,{\Lidy}+{8}\,{\Lidz}-{8}\,{\lnx}\,{\Licy}+{}\Biggl ({}-{6}\,{\lnx}+{4}\,{\lny}+{4}\Biggr ){}\,{\Licx} \nonumber \\
&& +{}\Biggl ({}-{4}\,{\lnx}+{2}\,{\lnx^2}+{2}\,{\pi^2}\Biggr ){}\,{\Libx}+{4}\,{\lny}\,{\lnx}\,{\Liby}+{13\over 24}\,{\lnx^4}+{}\Biggl ({}-{5\over 12}-{2\over 3}\,{\lny}\Biggr ){}\,{\lnx^3} \nonumber \\
&& +{}\Biggl ({}-{9}+{2}\,{\lny^2}-{7\over 6}\,{\pi^2}-{1\over 2}\,{\lny}\Biggr ){}\,{\lnx^2}+{}\Biggl ({}\Biggl ({2}+{4\over 3}\,{\pi^2}\Biggr ){}\,{\lny}-{3\over 2}\,{\pi^2}+{6}\,{\zeta_3}\Biggr ){}\,{\lnx} \nonumber \\
&& -{4}\,{\zeta_3}-{1\over 3}\,{\pi^4}-{2}\,{\lny^2}-{4}\,{\lny}\,{\zeta_3}-{\pi^2},
\end{eqnarray}
\begin{eqnarray}
D_{st}&=&{ }\Biggl [{2\over 3}\,{\Licx}-{2\over 3}\,{\lnx}\,{\Libx}+{44\over 9}\,{\Ls^2} \nonumber \\
&& +{}\Biggl ({}-{2\over 3}\,{\lnx^2}+{}\Biggl ({4\over 3}\,{\lny}+{26\over 9}\Biggr ){}\,{\lnx}+{4}\,{\lny}-{4\over 3}\,{\lny^2}-{389\over 27}-{2\over 3}\,{\pi^2}\Biggr ){}\,{\Ls} \nonumber \\
&& -{5\over 9}\,{\lnx^3}+{}\Biggl ({2\over 3}\,{\lny}+{37\over 18}\Biggr ){}\,{\lnx^2}+{}\Biggl ({}-{13\over 18}\,{\pi^2}-{11\over 9}\,{\lny}-{40\over 9}\Biggr ){}\,{\lnx}-{4\over 9}\,{\lny^3}+{29\over 9}\,{\lny^2} \nonumber \\
&& +{}\Biggl ({}-{11\over 9}\,{\pi^2}-{149\over 27}\Biggr ){}\,{\lny}-{2\over 9}\,{\pi^2}+{43\over 9}\,{\zeta_3}+{455\over 27}\Biggr ]{}\,{\ust},
\end{eqnarray}
\begin{eqnarray}
\label{eq:est}
E_{st}&=&{ }\Biggl [{2}\,{\Licx}-{2}\,{\lnx}\,{\Libx}+{}\Biggl ({1\over 3}\,{\lnx^2}+{}\Biggl ({4\over 3}\,{\lny}-{4\over 3}\Biggr ){}\,{\lnx}-{2\over 3}\,{\pi^2}-{4\over 3}\,{\lny^2}+{4}\,{\lny}-{29\over 3}\Biggr ){}\,{\Ls} \nonumber \\
&& +{1\over 9}\,{\lnx^3}-{19\over 18}\,{\lnx^2}+{}\Biggl ({}-{11\over 9}\,{\lny}+{1\over 3}\,{\pi^2}-{43\over 9}\Biggr ){}\,{\lnx}-{4\over 9}\,{\lny^3}+{29\over 9}\,{\lny^2}+{}\Biggl ({}-{14\over 9}\,{\pi^2}-{11\over 3}\Biggr ){}\,{\lny} \nonumber \\
&& +{29\over 9}\,{\zeta_3}+{1370\over 81}+{22\over 9}\,{\pi^2}\Biggr ]{}\,{\ust} \nonumber \\
&& +{}\Biggl [{4\over 3}\,{\lnx}\,{\Ls}+{1\over 9}\,{\lnx^3}+{2\over 3}\,{\lnx^2}+{}\Biggl ({}-{2\over 9}\,{\pi^2}-{32\over 9}\Biggr ){}\,{\lnx}+{2\over 3}\,{\pi^2}\Biggr ]{}\,{\ut} \nonumber \\
&& -{1\over 3}\,{\lnx^3}+{16\over 9}\,{\lnx^2}-{2\over 3}\,{\lnx}\,{\pi^2}-{2\over 3}\,{\lnx^2}\,{\Ls},
\end{eqnarray}
\begin{eqnarray}
\label{eq:fst}
F_{st}&=&{ }\Biggl [{}-{4\over 9}\,{\Ls^2}+{}\Biggl ({40\over 27}-{4\over 9}\,{\lnx}\Biggr ){}\,{\Ls}+{2\over 9}\,{\pi^2}-{100\over 81}+{20\over 27}\,{\lnx}-{2\over 9}\,{\lnx^2}\Biggr ]{}\,{\ust}.
\end{eqnarray}
Some of these results overlap with the analytic
expressions presented in Ref.~\cite{BDG} for the QED process
$e^+e^-\to e^+e^-$.
To obtain the QED limit from a QCD calculation
corresponds to setting $\CA = 0$, $\CF = 1$, $T_R
= 1$ as well as setting the cubic Casimir $C_3 = (N^2-1)(N^2-2)/N^2$ to $0$.
This means that we can directly compare $E_{st} (\propto \CF T_R\NF)$ 
and $F_{st} (\propto T_R^2\NF^2)$ but {\em not} $C_{st}$ which receives
contributions from both $C_3$ and $\CF^2$.
We see that (\ref{eq:est}) 
and (\ref{eq:fst}) agree with  Eqs.~(2.50) and (2.51) of~\cite{BDG} 
respectively. 
 
The other coefficients, $A_{st}$, $B_{st}$, $C_{st}$ and $D_{st}$ are new results.

\subsubsection{The process $q + q \to  q + q$}
\label{subsec:utex}
The mixed $t$- and $u$-channel process,
$q + q \to  q + q$
is fixed by $\B^8(t,s,u)$. 
We find that the finite two-loop contribution 
is given by Eq.~(\ref{eq:zi}) with 
\begin{eqnarray}
A_{ut}&=&{ }\Biggl [{2}\,{\Lidz}-{2}\,{\Lidx}-{2}\,{\Lidy}+{}\Biggl ({2}\,{\lnx}+{2}\,{\lny}+{23\over 3}\Biggr ){}\,{\Licx}+{}\Biggl ({4}\,{\lny}+{59\over 3}\Biggr ){}\,{\Licy} \nonumber \\
&& +{}\Biggl ({}-{2}\,{\lnx^2}+{}\Biggl ({}-{23\over 3}+{2}\,{\lny}\Biggr ){}\,{\lnx}+{1\over 3}\,{\pi^2}+{59\over 3}\,{\lny}\Biggr ){}\,{\Libx}+{}\Biggl ({2}\,{\lnx}\,{\lny}-{2}\,{\lny^2}\Biggr ){}\,{\Liby} \nonumber \\
&&-{121\over 9}\,{\Ls^2}+{}\Biggl ({11\over 3}\,{\lnx^2}+{592\over 27}-{22\over 9}\,{\lny}-{22\over 9}\,{\lnx}+{11\over 3}\,{\lny^2}\Biggr ){}\,{\Ls}-{1\over 6}\,{\lnx^4}+{}\Biggl ({}-{4\over 3}\,{\lny}+{14\over 9}\Biggr ){}\,{\lnx^3} \nonumber \\
&& +{}\Biggl ({}-{25\over 12}\,{\lny}-{31\over 6}+{7\over 12}\,{\pi^2}+{7\over 2}\,{\lny^2}\Biggr ){}\,{\lnx^2}-{1\over 3}\,{\lny^4}+{113\over 36}\,{\lny^3}+{}\Biggl ({}-{8\over 3}+{7\over 12}\,{\pi^2}\Biggr ){}\,{\lny^2}\nonumber \\
&& +{}\Biggl ({}-{2\over 3}\,{\lny^3}+{77\over 6}\,{\lny^2}+{}\Biggl ({7}-{17\over 6}\,{\pi^2}\Biggr ){}\,{\lny} +{695\over 216}+{59\over 36}\,{\pi^2}-{8}\,{\zeta_3}\Biggr ){}\,{\lnx} \nonumber \\
&& +{}\Biggl ({695\over 216}-{8}\,{\zeta_3}-{73\over 18}\,{\pi^2}\Biggr ){}\,{\lny}-{23213\over 1296}+{17\over 24}\,{\pi^4}-{485\over 18}\,{\zeta_3}+{73\over 18}\,{\pi^2}\Biggr ]{}\,{\sut} \nonumber \\
&& +{}\Biggl [{4}\,{\Lidz}+{}\Biggl ({}-{2}\,{\lnx}+{2}\,{\lny}-{12}\Biggr ){}\,{\Licx}+{}\Biggl ({2}\,{\lny}+{12}-{2}\,{\lnx}\Biggr ){}\,{\Licy} \nonumber \\
&& +{}\Biggl ({}\Biggl ({2}\,{\lny}+{12}\Biggr ){}\,{\lnx}+{2\over 3}\,{\pi^2}+{12}\,{\lny}\Biggr ){}\,{\Libx}+{2}\,{\lny}\,{\lnx}\,{\Liby}+{1\over 4}\,{\lnx^4}+{}\Biggl ({}-{\lny}-{19\over 12}\Biggr ){}\,{\lnx^3} \nonumber \\
&& +{}\Biggl ({}-{5\over 2}+{1\over 6}\,{\pi^2}+{5\over 2}\,{\lny^2}+{19\over 4}\,{\lny}\Biggr ){}\,{\lnx^2}+{}\Biggl ({29\over 4}\,{\lny^2}-{1\over 3}\,{\lny}\,{\pi^2}+{29\over 12}\,{\pi^2}+{1\over 3}\,{\lny^3}\Biggr ){}\,{\lnx} \nonumber \\
&& -{1\over 12}\,{\lny^4}+{19\over 12}\,{\lny^3}+{}\Biggl ({1\over 6}\,{\pi^2}+{5\over 2}\Biggr ){}\,{\lny^2}-{53\over 12}\,{\lny}\,{\pi^2}-{1\over 60}\,{\pi^4}\Biggr ]{}\,{\st}+{}\Biggl [{3}\,{\lnx^2}\Biggr ]{}\,{\tou}+{}\Biggl [{3}\,{\lny^2}\Biggr ]{}\,{\uot} \nonumber \\
&& +{32}\,{\Lidx}+{32}\,{\Lidy}+{}\Biggl ({}-{18}\,{\lnx}-{10}\,{\lny}-{2}\Biggr ){}\,{\Licx}+{}\Biggl ({}-{2}-{10}\,{\lnx}-{18}\,{\lny}\Biggr ){}\,{\Licy} \nonumber \\
&& +{}\Biggl ({2}\,{\lnx^2}+{}\Biggl ({2}-{10}\,{\lny}\Biggr ){}\,{\lnx}-{2}\,{\lny}\Biggr ){}\,{\Libx}+{}\Biggl ({}-{10}\,{\lnx}\,{\lny}+{2}\,{\lny^2}\Biggr ){}\,{\Liby}\nonumber \\
&& +{5\over 12}\,{\lnx^4} +{}\Biggl ({}-{\lny}-{7\over 12}\Biggr ){}\,{\lnx^3}+{}\Biggl ({2}-{27\over 2}\,{\lny^2}+{5\over 4}\,{\lny}-{4\over 3}\,{\pi^2}\Biggr ){}\,{\lnx^2} \nonumber \\
&& +{}\Biggl ({}-{\lny^3}-{3\over 4}\,{\lny^2}+{}\Biggl ({}-{10}+{6}\,{\pi^2}\Biggr ){}\,{\lny}-{1\over 4}\,{\pi^2}\Biggr ){}\,{\lnx}+{5\over 12}\,{\lny^4}-{7\over 12}\,{\lny^3}+{}\Biggl ({2}-{4\over 3}\,{\pi^2}\Biggr ){}\,{\lny^2} \nonumber \\
&& +{1\over 12}\,{\lny}\,{\pi^2}-{391\over 180}\,{\pi^4}+{5}\,{\pi^2},
\end{eqnarray}
\begin{eqnarray}
B_{ut}&=&{ }\Biggl [{3}\,{\Lidz}+{8}\,{\Lidx}+{8}\,{\Lidy}+{}\Biggl ({}-{12}\,{\lnx}+{6}\Biggr ){}\,{\Licx}+{}\Biggl ({6}-{8}\,{\lny}-{4}\,{\lnx}\Biggr ){}\,{\Licy} \nonumber \\
&& +{}\Biggl ({13\over 2}\,{\lnx^2}+{}\Biggl ({}-{6}-{\lny}\Biggr ){}\,{\lnx}+{6}\,{\lny}+{1\over 2}\,{\pi^2}\Biggr ){}\,{\Libx}+{11\over 2}\,{\lny^2}\,{\Liby} \nonumber \\
&& +{}\Biggl ({}-{11\over 6}\,{\lnx^2}+{}\Biggl ({11}\,{\lny}+{22\over 3}\Biggr ){}\,{\lnx}-{11\over 6}\,{\lny^2}+{176\over 3}+{44\over 3}\,{\lny}-{11\over 2}\,{\pi^2}\Biggr ){}\,{\Ls}+{1\over 6}\,{\lnx^4} \nonumber \\
&& +{}\Biggl ({}-{7\over 9}+{2\over 3}\,{\lny}\Biggr ){}\,{\lnx^3}+{}\Biggl ({}-{3}\,{\lny}+{4}\,{\pi^2}-{19\over 36}-{1\over 4}\,{\lny^2}\Biggr ){}\,{\lnx^2}+{1\over 12}\,{\lny^4} \nonumber \\
&& +{}\Biggl ({\lny^3}+{13\over 6}\,{\lny^2}+{}\Biggl ({}-{39\over 2}-{1\over 3}\,{\pi^2}\Biggr ){}\,{\lny}+{181\over 9}+{53\over 6}\,{\pi^2}\Biggr ){}\,{\lnx}-{1\over 2}\,{\lny^3} \nonumber \\
&& +{}\Biggl ({545\over 36}+{4}\,{\pi^2}\Biggr ){}\,{\lny^2}+{}\Biggl ({53\over 9}+{76\over 9}\,{\pi^2}\Biggr ){}\,{\lny}-{161\over 120}\,{\pi^4}-{30659\over 324}+{473\over 18}\,{\zeta_3}+{113\over 4}\,{\pi^2}\Biggr ]{}\,{\sut} \nonumber \\
&& +{}\Biggl [{6}\,{\Lidz}+{}\Biggl ({4}\,{\lny}-{4}\,{\lnx}\Biggr ){}\,{\Licx}+{}\Biggl ({4}\,{\lny}-{4}\,{\lnx}\Biggr ){}\,{\Licy} \nonumber \\
&& +{}\Biggl ({\lnx^2}+{\pi^2}+{2}\,{\lnx}\,{\lny}\Biggr ){}\,{\Libx}+{}\Biggl ({}-{\lny^2}+{4}\,{\lnx}\,{\lny}\Biggr ){}\,{\Liby}+{}\Biggl ({22\over 3}\,{\lny}-{22\over 3}\,{\lnx}\Biggr ){}\,{\Ls} \nonumber \\
&& +{5\over 24}\,{\lnx^4}+{}\Biggl ({}-{5\over 18}-{5\over 6}\,{\lny}\Biggr ){}\,{\lnx^3}+{}\Biggl ({}-{47\over 3}+{5\over 6}\,{\lny}+{1\over 4}\,{\pi^2}+{15\over 4}\,{\lny^2}\Biggr ){}\,{\lnx^2} \nonumber \\
&& +{}\Biggl ({128\over 9}-{1\over 6}\,{\lny^3}-{11\over 18}\,{\pi^2}-{1\over 2}\,{\lny}\,{\pi^2}-{5\over 6}\,{\lny^2}\Biggr ){}\,{\lnx}+{1\over 24}\,{\lny^4}+{5\over 18}\,{\lny^3}+{}\Biggl ({47\over 3}+{1\over 4}\,{\pi^2}\Biggr ){}\,{\lny^2} \nonumber \\
&& +{}\Biggl ({}-{128\over 9}+{11\over 18}\,{\pi^2}\Biggr ){}\,{\lny}-{1\over 40}\,{\pi^4}\Biggr ]{}\,{\st}+{}\Biggl [{4}\,{\lnx^2}\Biggr ]{}\,{\tou}+{}\Biggl [{4}\,{\lny^2}\Biggr ]{}\,{\uot} \nonumber \\
&& -{16}\,{\Lidx}-{16}\,{\Lidy}+{}\Biggl ({12}\,{\lnx}-{4}\,{\lny}-{2}\Biggr ){}\,{\Licx}+{}\Biggl ({}-{4}\,{\lnx}-{2}+{12}\,{\lny}\Biggr ){}\,{\Licy} \nonumber \\
&& +{}\Biggl ({}-{4}\,{\lnx^2}+{}\Biggl ({2}-{4}\,{\lny}\Biggr ){}\,{\lnx}-{2}\,{\lny}\Biggr ){}\,{\Libx}+{}\Biggl ({}-{4}\,{\lny^2}-{4}\,{\lnx}\,{\lny}\Biggr ){}\,{\Liby} \nonumber \\
&& +{}\Biggl ({11\over 3}\,{\lny^2}+{11\over 3}\,{\lnx^2}+{11\over 3}\,{\pi^2}-{22\over 3}\,{\lnx}\,{\lny}\Biggr ){}\,{\Ls}-{1\over 24}\,{\lnx^4}+{}\Biggl ({}-{7\over 6}\,{\lny}+{2}\Biggr ){}\,{\lnx^3} \nonumber \\
&& +{}\Biggl ({5\over 12}\,{\pi^2}-{163\over 9}-{17\over 4}\,{\lny^2}-{4\over 3}\,{\lny}\Biggr ){}\,{\lnx^2}+{}\Biggl ({}-{7\over 6}\,{\lny^3}-{10\over 3}\,{\lny^2}+{}\Biggl ({1\over 2}\,{\pi^2}+{254\over 9}\Biggr ){}\,{\lny}+{7\over 3}\,{\pi^2}\Biggr ){}\,{\lnx} \nonumber \\
&& -{1\over 24}\,{\lny^4}+{2}\,{\lny^3}+{}\Biggl ({5\over 12}\,{\pi^2}-{163\over 9}\Biggr ){}\,{\lny^2}+{8\over 3}\,{\lny}\,{\pi^2}+{421\over 360}\,{\pi^4}-{127\over 9}\,{\pi^2},
\end{eqnarray}
\begin{eqnarray}
C_{ut}&=&{ }\Biggl [{5}\,{\Lidz}+{2}\,{\Lidx}+{2}\,{\Lidy}+{}\Biggl ({2}\,{\lny}-{6}\,{\lnx}-{1}\Biggr ){}\,{\Licx}+{}\Biggl ({}-{4}\,{\lnx}-{1}\Biggr ){}\,{\Licy} \nonumber \\
&& +{}\Biggl ({5\over 2}\,{\lnx^2}+{}\Biggl ({1}+{\lny}\Biggr ){}\,{\lnx}-{\lny}+{5\over 6}\,{\pi^2}\Biggr ){}\,{\Libx}+{}\Biggl ({2}\,{\lnx}\,{\lny}+{3\over 2}\,{\lny^2}\Biggr ){}\,{\Liby} \nonumber \\
&& +{1\over 6}\,{\lnx^4}+{5\over 12}\,{\lnx^3}+{}\Biggl ({7\over 4}\,{\pi^2}-{1\over 2}\,{\lny}+{5\over 4}+{5\over 4}\,{\lny^2}\Biggr ){}\,{\lnx^2}-{1\over 12}\,{\lny^4}+{3\over 2}\,{\lny^3} \nonumber \\
&& +{}\Biggl ({\lny^3}-{19\over 4}\,{\lny^2}+{}\Biggl ({}-{5\over 6}\,{\pi^2}-{31\over 2}\Biggr ){}\,{\lny}+{41\over 12}\,{\pi^2}+{6}\,{\zeta_3}-{45\over 8}\Biggr ){}\,{\lnx}+{}\Biggl ({7\over 4}\,{\pi^2}+{15\over 4}\Biggr ){}\,{\lny^2} \nonumber \\
&& +{}\Biggl ({}-{141\over 8}+{6}\,{\zeta_3}+{13\over 3}\,{\pi^2}\Biggr ){}\,{\lny}-{511\over 16}+{20}\,{\zeta_3}+{109\over 12}\,{\pi^2}-{49\over 90}\,{\pi^4}\Biggr ]{}\,{\sut} \nonumber \\
&& +{}\Biggl [{10}\,{\Lidz}+{}\Biggl ({}-{6}\,{\lnx}+{6}\,{\lny}\Biggr ){}\,{\Licx}+{}\Biggl ({}-{6}\,{\lnx}+{6}\,{\lny}\Biggr ){}\,{\Licy} \nonumber \\
&& +{}\Biggl ({\lnx^2}+{4}\,{\lnx}\,{\lny}+{5\over 3}\,{\pi^2}\Biggr ){}\,{\Libx}+{}\Biggl ({6}\,{\lnx}\,{\lny}-{\lny^2}\Biggr ){}\,{\Liby}+{11\over 24}\,{\lnx^4}+{}\Biggl ({}-{11\over 6}\,{\lny}-{13\over 12}\Biggr ){}\,{\lnx^3} \nonumber \\
&& +{}\Biggl ({25\over 4}\,{\lny^2}+{13\over 4}\,{\lny}+{5\over 12}\,{\pi^2}-{5\over 2}\Biggr ){}\,{\lnx^2}+{}\Biggl ({}-{3\over 4}\,{\pi^2}+{12}+{1\over 6}\,{\lny^3}-{13\over 4}\,{\lny^2}-{5\over 6}\,{\lny}\,{\pi^2}\Biggr ){}\,{\lnx} \nonumber \\
&& -{1\over 24}\,{\lny^4}+{13\over 12}\,{\lny^3}+{}\Biggl ({5\over 2}+{5\over 12}\,{\pi^2}\Biggr ){}\,{\lny^2}+{}\Biggl ({}-{12}+{3\over 4}\,{\pi^2}\Biggr ){}\,{\lny}-{1\over 24}\,{\pi^4}\Biggr ]{}\,{\st}+{}\Biggl [{\lnx^2}\Biggr ]{}\,{\tou}\nonumber \\
&& +{}\Biggl [{\lny^2}\Biggr ]{}\,{\uot} -{8}\,{\Lidx}-{8}\,{\Lidy}+{}\Biggl ({6}\,{\lnx}-{2}\,{\lny}-{4}\Biggr ){}\,{\Licx}+{}\Biggl ({}-{4}-{2}\,{\lnx}+{6}\,{\lny}\Biggr ){}\,{\Licy} \nonumber \\
&& +{}\Biggl ({}-{2}\,{\lnx^2}+{}\Biggl ({4}-{2}\,{\lny}\Biggr ){}\,{\lnx}-{4}\,{\lny}\Biggr ){}\,{\Libx}+{}\Biggl ({}-{2}\,{\lnx}\,{\lny}-{2}\,{\lny^2}\Biggr ){}\,{\Liby} \nonumber \\
&& +{5\over 24}\,{\lnx^4}+{}\Biggl ({}-{3\over 2}\,{\lny}-{5\over 12}\Biggr ){}\,{\lnx^3}+{}\Biggl ({}-{9}-{3\over 4}\,{\lny^2}+{7\over 4}\,{\lny}+{5\over 12}\,{\pi^2}\Biggr ){}\,{\lnx^2} \nonumber \\
&& +{}\Biggl ({}-{3\over 2}\,{\lny^3}-{9\over 4}\,{\lny^2}+{}\Biggl ({16}-{1\over 6}\,{\pi^2}\Biggr ){}\,{\lny}+{1\over 4}\,{\pi^2}\Biggr ){}\,{\lnx}+{5\over 24}\,{\lny^4}-{5\over 12}\,{\lny^3}+{}\Biggl ({}-{9}+{5\over 12}\,{\pi^2}\Biggr ){}\,{\lny^2} \nonumber \\
&& +{11\over 12}\,{\lny}\,{\pi^2}+{203\over 360}\,{\pi^4}-{8}\,{\pi^2},
\end{eqnarray}
\begin{eqnarray}
D_{ut}&=&{ }\Biggl [{}-{2\over 3}\,{\Licx}-{2\over 3}\,{\Licy}+{}\Biggl ({}-{2\over 3}\,{\lny}+{2\over 3}\,{\lnx}\Biggr ){}\,{\Libx}+{44\over 9}\,{\Ls^2} \nonumber \\
&& +{}\Biggl ({26\over 9}\,{\lnx}-{2\over 3}\,{\lnx^2}+{26\over 9}\,{\lny}-{389\over 27}-{2\over 3}\,{\lny^2}\Biggr ){}\,{\Ls}-{5\over 9}\,{\lnx^3}+{}\Biggl ({37\over 18}+{1\over 3}\,{\lny}\Biggr ){}\,{\lnx^2} \nonumber \\
&& +{}\Biggl ({}-{7\over 18}\,{\pi^2}-{1\over 3}\,{\lny^2}-{40\over 9}\Biggr ){}\,{\lnx}-{5\over 9}\,{\lny^3}+{37\over 18}\,{\lny^2}+{}\Biggl ({}-{40\over 9}-{5\over 18}\,{\pi^2}\Biggr ){}\,{\lny} \nonumber \\
&& +{49\over 9}\,{\zeta_3}-{25\over 18}\,{\pi^2}+{455\over 27}\Biggr ]{}\,{\sut},
\end{eqnarray}
\begin{eqnarray}
\label{eq:eut}
E_{ut}&=&{ }\Biggl [{}-{2}\,{\Licx}-{2}\,{\Licy}+{}\Biggl ({}-{2}\,{\lny}+{2}\,{\lnx}\Biggr ){}\,{\Libx}+{1\over 9}\,{\lnx^3}-{19\over 18}\,{\lnx^2} \nonumber \\
&& +{}\Biggl ({1\over 3}\,{\lnx^2}+{}\Biggl ({}-{2}\,{\lny}-{4\over 3}\Biggr ){}\,{\lnx}-{29\over 3}+{1\over 3}\,{\lny^2}+{\pi^2}-{8\over 3}\,{\lny}\Biggr ){}\,{\Ls} \nonumber \\
&& +{}\Biggl ({}-{5\over 3}\,{\lny^2}+{2}\,{\lny}+{2\over 3}\,{\pi^2}-{43\over 9}\Biggr ){}\,{\lnx}-{31\over 18}\,{\lny^2}+{}\Biggl ({8\over 9}\,{\pi^2}-{11\over 9}\Biggr ){}\,{\lny} \nonumber \\
&& +{1370\over 81}-{5\over 2}\,{\pi^2}+{47\over 9}\,{\zeta_3}\Biggr ]{}\,{\sut} \nonumber \\
&& +{}\Biggl [{}\Biggl ({4\over 3}\,{\lnx}-{4\over 3}\,{\lny}\Biggr ){}\,{\Ls}+{1\over 9}\,{\lnx^3}+{}\Biggl ({}-{1\over 3}\,{\lny}+{2\over 3}\Biggr ){}\,{\lnx^2}+{}\Biggl ({1\over 3}\,{\lny^2}+{1\over 9}\,{\pi^2}-{32\over 9}\Biggr ){}\,{\lnx} \nonumber \\
&& -{1\over 9}\,{\lny^3}-{2\over 3}\,{\lny^2}+{}\Biggl ({}-{1\over 9}\,{\pi^2}+{32\over 9}\Biggr ){}\,{\lny}\Biggr ]{}\,{\st} \nonumber \\
&& +{}\Biggl ({}-{2\over 3}\,{\lny^2}-{2\over 3}\,{\lnx^2}-{2\over 3}\,{\pi^2}+{4\over 3}\,{\lnx}\,{\lny}\Biggr ){}\,{\Ls}-{1\over 3}\,{\lnx^3}+{}\Biggl ({1\over 3}\,{\lny}+{16\over 9}\Biggr ){}\,{\lnx^2} \nonumber \\
&& +{}\Biggl ({}-{32\over 9}\,{\lny}+{1\over 3}\,{\lny^2}-{1\over 3}\,{\pi^2}\Biggr ){}\,{\lnx}+{16\over 9}\,{\lny^2}-{1\over 3}\,{\lny^3}+{16\over 9}\,{\pi^2}-{1\over 3}\,{\lny}\,{\pi^2},
\end{eqnarray}
\begin{eqnarray}
\label{eq:fut}
F_{ut}&=&{ }\Biggl [{}-{4\over 9}\,{\Ls^2}+{}\Biggl ({}-{4\over
9}\,{\lny}+{40\over 27}-{4\over 9}\,{\lnx}\Biggr ){}\,{\Ls}+{20\over
27}\,{\lnx}-{2\over 9}\,{\lnx^2}+{20\over 27}\,{\lny}-{2\over
9}\,{\lny^2}-{100\over 81}\Biggr ]{}\,{\sut}.\nonumber \\
\end{eqnarray}
As in Section~\ref{subsec:stex},
we can compare
some of these results with the analytic
expressions presented in Ref.~\cite{BDG} for the QED process
$e^+e^-\to e^+e^-$,
and we see that (\ref{eq:eut}) 
and (\ref{eq:fut}) agree with  Eqs.~(2.55) and (2.56) of~\cite{BDG} 
respectively. 
 
The other coefficients, $A_{ut}$, $B_{ut}$, $C_{ut}$ and $D_{ut}$ 
represent new results.
 
\section{Summary}
\label{sec:conc}

In this paper we discussed the two-loop QCD corrections to the scattering of
two identical massless quarks.   For the annihilation process, both $s$-channel
and $t$-channel graphs are present.   The interference of $s$-channel tree and
two-loop graphs is determined by $\A^{8 (2 \times 0)}(s,t,u)$ which  is the
same function that describes distinct quark scattering in the $s$-channel.   
Similarly, the interference of the $t$-channel tree and two-loop graphs is
fixed by $\A^{8 (2 \times 0)}(t,s,u)$.  Explicit expressions for $\A^{8 (2
\times 0)}$ are given in~\cite{qqQQ}. The modification to the matrix elements
due to the interference of the $s$-channel tree graph with the $t$-channel
two-loop graphs (and vice versa) is represented by $\B^{8 (2 \times
0)}(s,t,u)$ (see Eq.~(\ref{eq:AB})).    To obtain $\B^{8 (2 \times 0)}$,  
we have used conventional
dimensional regularisation and the \MSbar\  renormalisation scheme to
compute the interference of the tree and two-loop graphs summed over spins and
colours.  

The pole structure for $\B^{8 (2 \times 0)}(s,t,u)$ is given in
Eq.~(\ref{eq:poles}) while expressions for the finite parts are given for  the
mixed $s$- and  $t$-channels and mixed $u$- and $t$-channels  in
Secs.~\ref{subsec:stex} and \ref{subsec:utex} respectively.  Together with the
analogous expressions for unlike-quark scattering given in
Ref.~\cite{qqQQ}, they complete the analytic formulae required to describe the
two-loop contribution to quark-quark scattering through to $\O{\ep^0}$.

These results form an important part of the next-to-next-to-leading order
predictions for jet cross sections in hadron-hadron collisions.    However,
they are only a part of the whole and must be combined with the  tree-level $2
\to 4$, the one-loop $2 \to 3$ as well as the square of the one-loop $2 \to 2$
processes to yield physical cross sections.    For the most part, the matrix
elements themselves are available in the literature.  Each of the contributions
is  divergent in the infrared limit and a systematic procedure for analytically
canceling the infrared divergences needs to be established for semi-inclusive 
jet cross sections.    Recent progresses in  determining the singular limits of
tree-level matrix elements when two particles are unresolved~\cite{tc,ds} and
the soft and collinear limits of one-loop amplitudes~\cite{sone,cone}, together
with the analytic cancellation of the infrared singularities in the somewhat
simpler case  of $e^+e^- \to {\rm photon} + {\rm jet}$ at next-to-leading order
\cite{aude}, suggest that the technical problems will soon be solved for
generic $2 \to 2$ scattering processes.   There are additional problems due to
initial state radiation.  However, the recent steps taken towards the
determination of the three-loop splitting functions~\cite{moms1,moms2,Gra1}
are also promising.    We therefore expect that the problem of the analytic
cancellation of the infrared divergences will soon be addressed thereby
enabling  the construction of numerical programs to provide
next-to-next-to-leading order QCD estimates of jet production in hadron
collisions.

\section*{Acknowledgements}

C.A. acknowledges the financial support of the Greek Government and
M.E.T. acknowledges financial support from CONACyT and the CVCP. 
We gratefully acknowledge the support of
the British Council and German Academic Exchange Service under ARC project
1050.  This work was supported in part by the EU Fourth Framework Programme
`Training and Mobility of Researchers', Network `Quantum Chromodynamics and
the Deep Structure of Elementary Particles', contract FMRX-CT98-0194
(DG-12-MIHT).

\end{document}